\documentstyle[twocolumn,aps,prl,psfig]{revtex}
\begin{document}
\draft
\twocolumn[\hsize\textwidth\columnwidth\hsize\csname @twocolumnfalse\endcsname

\title{High-field Electron Spin Resonance of Cu$_{1-x}$Zn$_{x}$GeO$_{3}$}

\author{A. K. Hassan$^1$, L. A. Pardi$^2$, G. B. Martins$^2$, G. Cao$^2$, L-C. Brunel$^1$}
\address{$^1$National High Magnetic Field Lab and Department of Physics, Florida
State University, Tallahassee,\\
Florida 32310,USA}
\address{$^2$National High Magnetic Field Lab, Tallahassee,\\
Florida 32310,USA}
\maketitle

\begin{abstract}
High-Field Electron Spin Resonance measurements were made on powder samples of
Cu$_{1-x}$Zn$_{x}$GeO$_{3}$ (x=0.00, 0.01, 0.02, 0.03 and 0.05) at
different frequencies (95, 110, 190, 220, 330 and 440 GHz) at low temperatures. 
The spectra of the doped samples show resonances whose positions are dependent on 
Zn concentration, frequency and temperature. The analysis of intensity variation 
of these lines with temperature allows us to identify them as originating 
in transitions within states situated {\it inside} the Spin Peierls gap. 
A qualitative explanation of 
the details of the spectra is possible if we assume that these states in the gap 
are associated with ``loose'' spins created near the Zn impurities, as recently 
theoreticaly predicted. A new phenomenon of quenching of the ESR signal 
across the Dimerized to Incommensurate phase-boundary is observed.
\end{abstract}


\vskip2pc] \narrowtext

Since the discovery that CuGeO$_3$ undergoes a Spin-Peierls (SP) transition
at $T_{SP}=14$ K \cite{hase1} this compound has been extensively investigated. 
The Cu$^{2+}$ ions ($S=1/2$) form chains along the $c$ axis
and, due to the coupling of these spins with the lattice vibrations, there is 
a dimerization of the chains below $T_{SP}$ causing the opening of a gap
($\Delta \approx 24$ K) in the spin-spectrum above a singlet ground-state.
Part of the interest in this material is associated with the
possibility of doping it, which was not feasible with the
previously known organic SP materials\cite{bloch}. The doping may be achieved
in two ways: (i) by the substitution of Cu$^{2+}$ by $S=0$ ions (Zn$%
^{2+}$) or $S>1/2$ ions (Ni$^{2+}$, Mn$^{2+}$)\cite{oseroff} (ii) by the
substitution of Si for Ge\cite{oseroff}. Some unexpected effects appeared
upon doping: (i) the gap closes for very small doping concentration (with
less than 1\% of Si doping the magnetic susceptibility shows almost no trace
of the gap\cite{poirier}); (ii) the antiferromagnetic (AF) fluctuations
increase upon doping and the system undergoes a transition to a N\'{e}el
phase at a temperature much higher than expected; (iii) a coexistence 
of antiferromagnetism and dimerization has been reported\cite{martin}.

The motivation for the high-field Electron Spin Resonance (ESR) measurements
presented in this letter came from the recent theoretical prediction that the
doping with non-magnetic impurities (Zn$^{2+}$ in the present work) should
introduce states inside the gap\cite{martins}.In the vicinity of each Zn
ion a spin $1/2$ state is predicted to be located. These states are 
weakly interacting and their collective behavior 
likely belong to the universality class of the 
Randon Antiferromagnetic Heisenberg chain\cite{martins}\cite{ma}. 
The weight inside the gap 
should grow like $x$ (for small $x$) and so one expects that the behavior 
of these states should
depend on doping concentration. Inelastic Neutron Scatering (INS) 
measurements did not report low
energy states\cite{martin}, but this is likely caused by the presence of strong 
elastic scattering. As these states
were supposed to be magnetic ($S\neq 0$)\cite{martins} ESR seems to be the
appropriate technique to observe them. Up to now (to our knowledge) the
majority of the literature on doped CuGeO$_3$ has been focussed on the AF or on the
incommensurate (IC) phases. Due to the the challenging experimental results 
described above and due to the possible relevance of the in-gap states 
to this issue, the work for this letter has concentrated on the narrow
temperature window $T_{N}(x)<T<T_{SP}(x)$.

Two sets of polycrystalline Cu$_{1-x}$Zn$_x$GeO$_3$ samples with $x=0.00,0.01,0.02,0.03$
and $0.05$ were grown by the usual solid state reaction.
The two sets were prepared with different grinding and heating times . ESR and
magnetic susceptibility measurements yielded the same results on both sets,
and the susceptibility data are in good agreement with those of Weiden
et al.\cite{weiden}. The same amount of marker -MnO diluted in MgO - was
later mixed in with one set of samples, to serve as a field marker and 
intensity reference\cite{burg}.
The ESR measurements on this compound were performed with the National High
Magnetic Field Laboratory high-field ESR spectrometer. Different frequencies
were investigated ($95, 110, 190, 220, 330$ and $440$ GHz) using Gunn diode
sources with oversized waveguides and a single pass transmission probe. A 17
Tesla superconducting magnet was used to sweep the
field. A detailed description of the setup will be given elsewhere\cite{brunel}.

The ESR spectrum of polycrystalline CuGeO$_{3}$ shows the typical pattern of
a powder distribution with rhombic $g$ anisotropy. The $g$ values obtained
at $20K$ are $g_{b}=2.256$, $g_{a}=2.154$ and $g_{c}=2.061$, which agree with values
reported in the literature\cite{honda}\cite{oseroff2}. As the temperature is decreased below 
T$_{SP}$ , the ESR lines broaden and
decrease in amplitude along the three directions $a$, $b$ and $c$. 
This behavior has already been observed at X-band\cite{honda}.

\begin{figure}[htbp]
\centerline{\psfig{figure=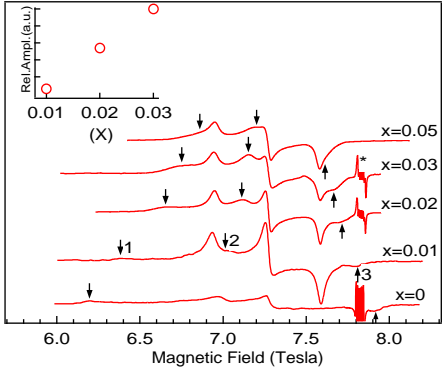,height=5cm}}
\vspace{0.5cm}
\caption{ESR spectra of Cu$_{1-x}$Zn$_x$GeO$_3$ for different Zn concentrations ($x$) at 
$f=220$ GHz and $T=4.2K$.
The extra features are marked by arrows and labeled 1, 2 and 3. 
The inset shows the amplitude of EF 1 at $4.2 K$ measured 
relative to the marker, as a function of $x$. The signal labeled by an * comes from the marker.}
\end{figure}

The ESR spectra of Cu$_{1-x}$Zn$_x$GeO$_3$ at $4.2K$ and $220$ GHz are shown
in Fig.1 for different values of Zn doping. Two different sets of resonances
are observed. The main features (MF) appear at transition fields coincident with
those of pure CuGeO$_3$ at higher temperatures. The extra features (EF)
marked in the figure by arrows and labeled as 1, 2, and 3 are specific to the
doped samples and represent the main focus of the present work. The resonance
fields, linewidths and amplitudes of these features depend on the Zn
concentration, frequency and temperature. This behavior
excludes the possibility that their origin could be a spurious
paramagnetic phase in the samples. Moreover, the two sets of samples studied
here gave the same results. It is clear in Fig.1 that the spectra of the $x=0.00$ sample 
shows also EF at low and high fields. We associate this to the fact that the sample 
has a residual susceptibility below $T_{SP}$, indicating that some intrinsic 
chain breaking mechanism is affecting its magnetic properties \cite{weiden}. This 
means that in reality the ``$x=0.0$'' sample should be considered as having an $x$ value 
between 0.00 and 0.01. The observed position of the EF in the nominally pure sample fits in 
the general trend of shift described below for a very low doping level.

It is also worth noting that the temperature behavior of the main features
depends on the Zn content. The intensity of these lines decreases more
rapidly with temperature for lower doping levels. 
In the case of the pure CuGeO$_{3}$ sample, the expression obtained by Bulaevskii\cite{bula} 
for an alternating chain of spins in the Hartree-Fock approximation, 
$I(T)=\frac{1}{T}\exp ^{-\frac{m}{T}}$, was used to analyse the temperature dependence of 
the ESR line intensity below $T_{SP}$, where $m$ is related to the SP gap $\Delta$ and to 
the exchange interaction $J$. Taking the value of $J=90 K$ \cite{hase1}, 
the observed value of the SP gap is $\Delta = 26 \pm 2 K$, in agreement with 
the literature \cite{nishi}\cite{oseroff2}. However, as expected, the extension 
of this analysis to the doped samples is not adequate\cite{weiden}. Therefore
taking into account the zero-field energy
gap $\Delta $, as well as the Zeeman term $g\mu _BBM_S$, the experimental data
presented in Fig.2(a)
fit quite well with theoretical predictions for the difference in population
between two energy levels based on Boltzmann statistics. 
For the doped samples,
the gap is estimated to be $23.5 \pm 1.5 K$, $22 \pm 1 K$ and $20 \pm 1 K$ 
for $x=0.01$, $%
0.02$ and $0.03$ respectively. These values compare well with the results of
neutron scattering\cite{martin}, 
showing that the original gap is partially supressed with Zn doping, 
in agreement with previous calculations\cite{martins}.

\begin{figure}[htbp]
\centerline{\psfig{figure=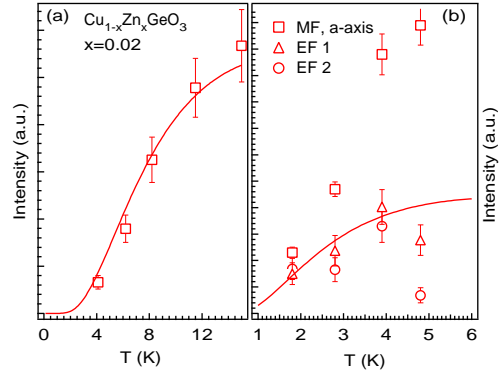,height=5cm}}
\vspace{0.5cm}
\caption{(a) Temperature dependence of the main feature along the $a$ axis (MF ``a") 
for $x=0.02$ at 220 GHz. (b) 
Temperature dependence of  MF ``a" and EF 1 and 2 at 188 GHz.
The solid lines are theoretical curves (see text).}
\end{figure}

Fig.2(b) shows the temperature dependence of the intensities of the main
feature along the $a$ axis and the EF 1 and 2 for the $x=0.02$
sample. In all Zn doped samples studied the Zn-induced features start to appear
only at temperatures between $5$ and $6K$. The temperature behavior of the 
intensity of the EF for $x=0.02$, as depicted in Fig.2(b), 
can be tentatively associated with
transitions within a band of S=1/2 states positioned at an energy $\delta$ 
above the ground state. 
Following the same procedure as for the main features in the doped samples, 
$\delta$ is estimated at $8 \pm 2 K$. Even 
though we had a limited temperature range to follow the EF, their intensity 
versus temperature behavior, if compared with that of the main features (see 
Fig.2(b)), can only be explained if associated with a resonance coming 
from states positioned {\it inside} the SP gap. The fact that the intensity 
of the EF sharply drops above $5 K$ (see Fig.2(b)) could indicate the 
presence of some relaxation mechanism.

Comparing the spectra obtained at the same frequency (220 GHz) and
temperature ($4.2$ K) for the different Zn concentrations (see Fig.1), the
resonance field positions of the in-gap states exhibit a dependence on the
Zn content. As x is increased from $0.01$ to $0.05$, the EF 1 and 2 shift
upwards in field toward the main features along the $b$ and $a$ axes
respectively, while EF 3 moves down in field toward the main feature along
the $c$ axis. The 
inset in Fig.1 also shows that as the doping level $x$ increases, the EF
become more intense.

\begin{figure}[htbp]
\centerline{\psfig{figure=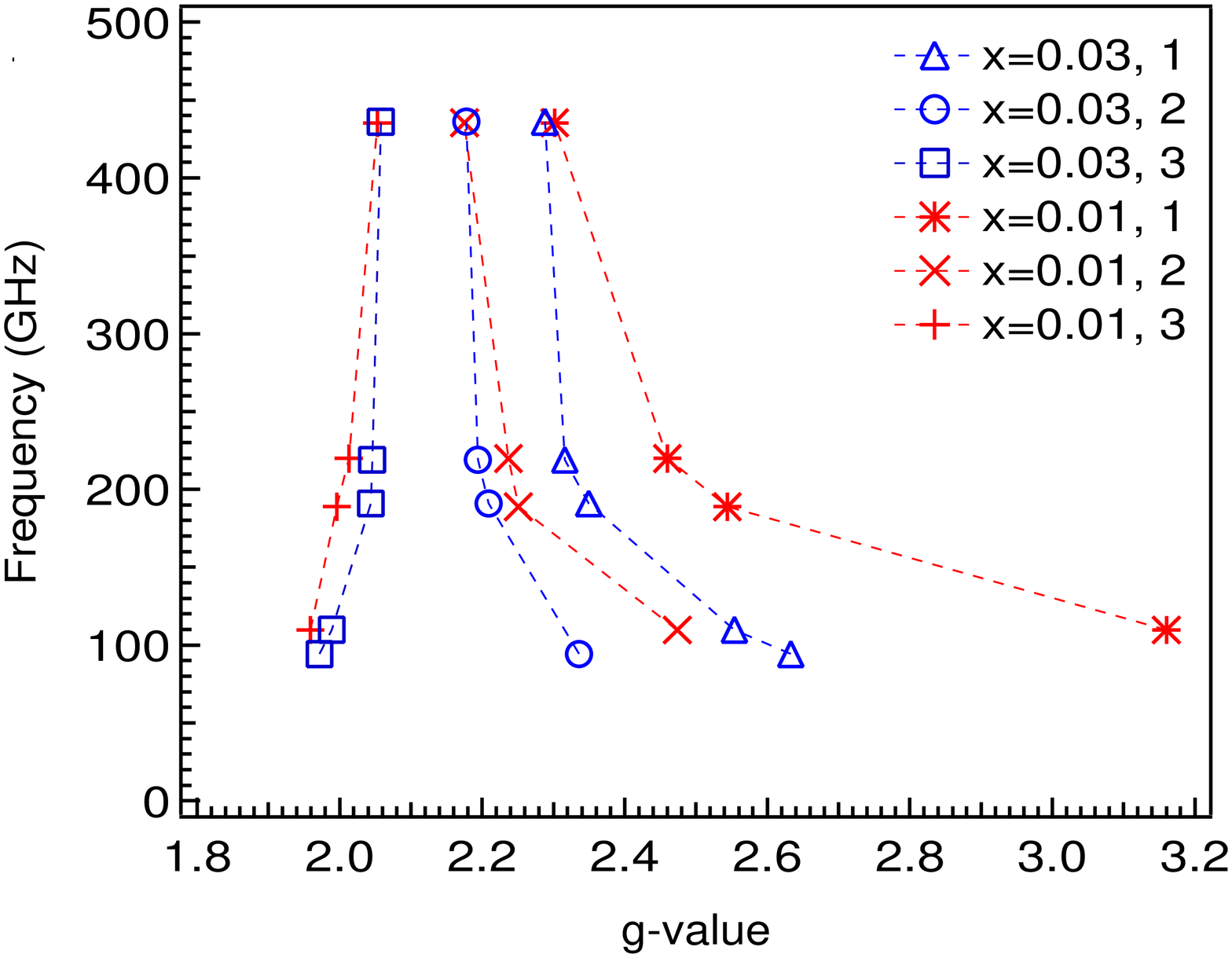,height=5cm}}
\vspace{0.5cm}
\caption{Frequency dependence of the g value of EF 1, 2 and 3 measured at 
$4.2K$ for $x=0.01$ and $x=0.03$. The $x=0.02$ sample exhibits similar 
behavior. Dashed lines are guides to the eye.}
\end{figure}

The frequency dependence of the EF has been studied keeping the temperature
and doping level fixed. The results are shown in
Fig.3 for two different doping levels. 
It is observed that the g-values of EF 1 and 2 decrease with
increasing frequency (starting with 95 GHz) while the g-value of EF 3
increases, and then they become frequency independent at higher frequencies. 
Moreover, these EF broaden and decrease in
amplitude as the frequency is lowered, that may explain why they were not
detected at lower frequencies\cite{hase2}, and again an indication that 
some relaxation mechanism is active.

Fig.4 shows the ESR spectra of Cu$_{1-x}$Zn$_x$GeO$_3$ for $x=0.01,0.02$ and 
$0.03$ measured at $4.2K$ and at different frequencies (220, 330 and 
440 GHz). Assuming the values obtained for the SP gap in the doped system, 
it can be seen that the frequency 440 GHz corresponds to resonance fields in the IC 
phase of the system, while 220 GHz , and therefore all lower frequencies, 
correspond to fields in the dimerized (D) phase. 
However, at 330 GHz (10
to 12 Tesla in field), the ESR signals from both Cu$_{1-x}$Zn$_x$GeO$_3$ and
the marker are almost completely quenched. As the temperature is increased, 
the signals
start to reappear and their intensity increases with temperature. This quenching
could be related to the phase transition from the dimerized to the
incommensurate phase which might cause damping of the microwaves. This
effect at 330 GHz is restricted to the doped samples only and has not been
observed in the pure compound around its D to IC phase transition. It is
possible that this transition is much broader in the doped material than in
the pure, as observed by Kiryukhin et al.\cite{kiryukhin}. This effect, 
which is beyond the scope of this work, is not well
understood at the moment and more experiments are being carried out to clarify it.

\begin{figure}[htbp]
\vspace{-0.5cm}
\centerline{\psfig{figure=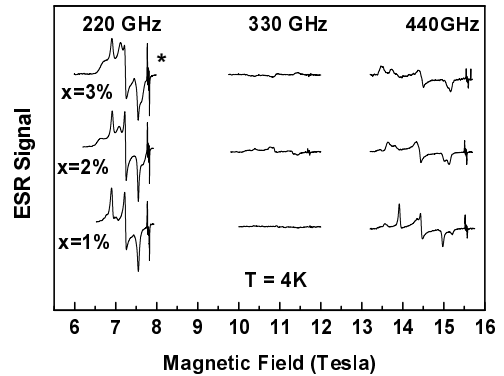,width=7.5cm}}
\vspace{-3.5cm}
\caption{ESR spectra of Cu$_{1-x}$Zn$_x$GeO$_3$ measured at $4.2K$ for $x=0.01$, $0.02$ and 
$0.03$, at the different frequencies $220$, $330$ and 440 GHz. The signal labeled by an * 
comes from the marker.}
\end{figure}

The temperature
dependence of the intensity of the central feature of the powder spectrum
(see Fig 2a), shows that the main effect of doping is to reduce the original gap, as
already observed with neutron scattering\cite{martin}. The appearence of low
temperature extra features in the ESR spectrum upon doping, 
and their coexistence with the
main features, show that there are two distinct ESR active entities in the
system. The main features are readily attributed to the undisturbed CuGeO$_3$
chains, the EF to in-gap states which can
be visualized as quasi-isolated spin doublets created in the antiferromagnetic
chains upon doping by non-magnetic impurities as predicted in 
ref. \cite{martins}.

\begin{figure}[htbp]
\centerline{\psfig{figure=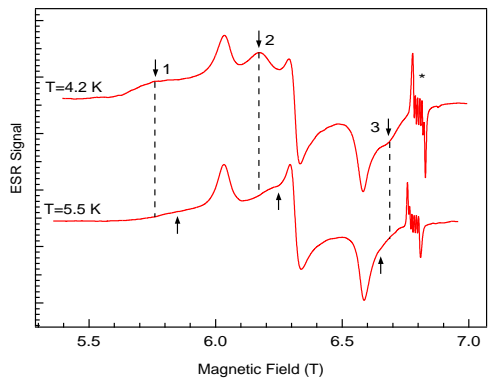,height=5cm}}
\vspace{0.5cm}
\caption{Temperature dependence of the resonance field of EF 1, 
2 and 3 for the $x=0.03$ sample 
at 191 GHz.}
\end{figure}

We first analyse the temperature dependence, since its explanation is related 
to the dependencies on Zn$^{2+}$ concentration and on frequency. The temperature dependence 
of the in-gap states resonance fields (Fig.5) is a common behavior in one 
dimensional spin systems 
\cite{gateschi}. It may be ascribed to the increase of short range order (SRO) at lower 
temperatures. Although there is no long range order in 1-D spin systems, the build up of 
some SRO is enough to create a local field (of dipolar origin) in the chains. As the 
correlation length increases at lower temperatures, the local field is stronger, 
causing a shift 
in the resonance field (see Fig.5). The presence of this correlation length dependent local field 
explains the fact that the in-gap states give resonance signals 
that, at low doping and low frequency, are substantialy shifted from the main features (see Fig.1). Martins 
et al. \cite{martins} assert that 
these states in the gap derive from the presence of ``loose" Cu$^{2+}$ ions that are created 
by the introduction of the Zn$^{2+}$ impurities. There is also evidence \cite{martins}\cite{martins2} 
that doping causes the ``loose" spins to feel a considerable increase in the AF 
correlations with the neighbors, 
i. e., there is an increase of SRO in the region where the ``loose" spin is located. As a consequence 
the ``loose" spin will feel a larger local field than the spins farther from the impurity, causing 
its resonance to be shifted in relation to the resonance field of the other spins, giving rise to 
an extra feature in the ESR spectrum. It is also possible to show that the positions of the three EF in the powder 
spectra are consistent with the expected orientation of this local field for an AF 1-D spin system.
Actually, an up-field shift is predicted for an external field paralell to the 
chain for an AF 1-D system, while down-field shifts are predicted in the 
perpendicular directions\cite{okamoto}. This prediction coherently explains 
the observed field positions of the EF compared to the main features, assuming that EF 3 corresponds 
to the chain direction, and EF 1 and 2 correspond to the perpendicular ones.
This enhanced local field near the Zn$^{2+}$ readily explains the frequency behavior displayed 
in Fig.3, because a constant local field should give a smaller g-shift at higher frequencies, i. 
e., at higher resonance fields.

To explain why the shift is smaller at higher doping (see Fig.1), a behavior that at first 
sight seems counterintuitive, we first note 
that a higher doping concentration does not enhance the increase in SRO near the Zn$^{2+}$ sites, it just 
makes it more frequent along the chain, without causing any increase in the local field felt by the 
``loose" spins. But, as the doping level increases, the ``loose" 
spins should start to interact with each other through an effective AF interaction \cite{martins}
\cite{hase3}, whose strength increases with doping level; this additional interaction will 
add some quantun fluctuations to the spin system, diminishing the local 
field felt by the ``loose" spins 
and consequently the shift in the resonance field. Currently a theoretical 
effort is being made to develop quantitatively 
these models, using numerical techniques\cite{martins3}.

In summary, the 
specific contribution of the multifrequency high-field 
ESR technique in this study resides in the fact
that i) it allows the observation of the in-gap states for the first time and
ii) it offers the possibility to have an insight into the physics of these
states {\it \ via} the observation of the principal values of the g tensor connected with
these excitations and their behavior with the doping level, 
temperature and frequency.

The authors acknowledge valuable discussions with E. Dagotto, W. Moulton and 
T. Brill, and thank S. McCall for susceptibility measurements and J. Crow for 
his support. This work was funded by an internal grant from the National 
High Magnetic Field Laboratory. L. A. P. thanks the Human Frontier Science 
Program for support (RG-349/94).

\end{document}